\documentclass[11pt,a4paper,english,nofootinbib]{revtex4}
\usepackage{lmodern}
\usepackage{lmodern}

\usepackage[T1]{fontenc}
\usepackage[latin9]{inputenc}
\usepackage{babel}
\usepackage{mathrsfs}
\usepackage{amsmath}
\usepackage{amssymb}
\usepackage{esint}
\usepackage[unicode=true,pdfusetitle,
 bookmarks=true,bookmarksnumbered=false,bookmarksopen=false,
 breaklinks=false,pdfborder={0 0 1},backref=false,colorlinks=false]
 {hyperref}

\makeatletter

\@ifundefined{textcolor}{}
{%
 \definecolor{BLACK}{gray}{0}
 \definecolor{WHITE}{gray}{1}
 \definecolor{RED}{rgb}{1,0,0}
 \definecolor{GREEN}{rgb}{0,1,0}
 \definecolor{BLUE}{rgb}{0,0,1}
 \definecolor{CYAN}{cmyk}{1,0,0,0}
 \definecolor{MAGENTA}{cmyk}{0,1,0,0}
 \definecolor{YELLOW}{cmyk}{0,0,1,0}
}

\usepackage{babel}
\usepackage{babel}
\usepackage{babel}
\usepackage{babel}
\usepackage{babel}
\usepackage{babel}
\usepackage{babel}
\usepackage{graphicx}
\def\b{\begin{equation}}
\def\e{\end{equation}}

\@ifundefined{textcolor}{}{%
 \definecolor{BLACK}{gray}{0}
 \definecolor{WHITE}{gray}{1}
 \definecolor{RED}{rgb}{1,0,0}
 \definecolor{GREEN}{rgb}{0,1,0}
 \definecolor{BLUE}{rgb}{0,0,1}
 \definecolor{CYAN}{cmyk}{1,0,0,0}
 \definecolor{MAGENTA}{cmyk}{0,1,0,0}
 \definecolor{YELLOW}{cmyk}{0,0,1,0}
 }

\usepackage{latexsym}\usepackage{bm}

\makeatother

\begin{document}
\title{{\normalsize{}{}New approach to conserved charges of generic gravity
in $AdS$}}
\author{{\normalsize{}{}Emel Altas}}
\email{emelaltas@kmu.edu.tr}

\affiliation{Department of Physics,\\
 Karamanoglu Mehmetbey University, 70100, Karaman, Turkey}
\author{{\normalsize{}{}Bayram Tekin}}
\email{btekin@metu.edu.tr}

\affiliation{Department of Physics,\\
 Middle East Technical University, 06800, Ankara, Turkey}
\date{{\normalsize{}{}\today}}

\begin{abstract}
Starting from a divergence-free rank-4 tensor
of which the trace is the cosmological Einstein tensor, we give a construction
of conserved charges in Einstein's gravity and its higher derivative
extensions for asymptotically anti-de Sitter spacetimes.
The current yielding the charge is explicitly gauge-invariant, and
the charge expression involves the linearized Riemann tensor at the
boundary. Hence, to compute the mass and angular momenta in these spacetimes,
one just needs to compute the linearized Riemann tensor. We give two examples.
\end{abstract}

\maketitle
\section{{\normalsize{}{}Introduction}}

Let us start with a seemingly innocent question which will have far-reaching
consequences for the conserved charges of gravity theories. Given
the Riemann tensor $R^{\nu}\thinspace_{\mu\beta\sigma}$, its single
trace (over the first and third indices) yields the Ricci tensor $R_{\mu\sigma}$;
is there a rank-4 tensor whose single trace is not the Ricci
tensor but the (cosmological) Einstein tensor, $\text{\ensuremath{{\cal {G}}}}_{\mu\sigma}=R_{\mu\sigma}-\frac{1}{2}Rg_{\mu\sigma}+\Lambda g_{\mu\sigma}$,
with the condition that this four-index tensor has the symmetries
of the Riemann tensor and it is divergence-free just like the Einstein
tensor? Remarkably the answer is affirmative: the tensor
\begin{equation}
\text{\ensuremath{{\cal {P}}}}^{\nu\mu\beta\sigma}:=R^{\nu\mu\beta\sigma}+g^{\sigma\nu}R^{\beta\mu}-g^{\beta\nu}R^{\sigma\mu}+g^{\beta\mu}R^{\sigma\nu}-g^{\sigma\mu}R^{\beta\nu}+\left(\frac{R}{2}-\frac{\Lambda\left(n-3)\right)}{n-1}\right)\left(g^{\beta\nu}g^{\sigma\mu}-g^{\sigma\nu}g^{\beta\mu}\right)\label{eq:Ptensor}
\end{equation}
whose construction will be given below does the job. It is divergence-free for all smooth metrics, \textit{i.e.} without the
use of any field equations 
\begin{equation}
\nabla_{\nu}{\cal {P}}^{\nu}\thinspace_{\mu\beta\sigma}=0,\label{divP}
\end{equation}
and its trace is the cosmological Einstein tensor as desired 
\begin{equation}
{\cal {P}}^{\nu}\thinspace_{\mu\nu\sigma}=(3-n)\text{\ensuremath{{\cal {G}}}}_{\mu\sigma}.
\end{equation}
Clearly, the interesting exception is that one cannot do this construction in three dimensions. What happens for $n=3$ is that the ${\cal {P}}$-tensor
vanishes \textit{identically} since, due to the vanishing
of the Weyl tensor, the Riemann and the Ricci tensors carry the same amount of information and the Riemann tensor can be expressed in terms of the Ricci tensor as 
\begin{equation}
R^{\nu\mu\beta\sigma}=R^{\nu\beta}g^{\mu\sigma}+R^{\mu\sigma}g^{\nu\beta}-R^{\nu\sigma}g^{\mu\beta}-R^{\mu\beta}g^{\nu\sigma}-\frac{R}{2}\left(g^{\nu\beta}g^{\mu\sigma}-g^{\mu\beta}g^{\nu\sigma}\right).
\end{equation}
Therefore, in some sense, the ${\cal {P}}$-tensor (\ref{eq:Ptensor})
is an obstruction for a smooth $generically$ curved metric to be
three dimensional. This can also be seen from the following identity: the Gauss-Bonnet combination $\chi_{\text{GB}}:=R^{\nu\mu\beta\sigma}R_{\nu\mu\beta\sigma}-4R_{\mu\nu}R^{\mu\nu}+R^{2}$
vanishes identically in three dimensions, and it is easy to show that the
contraction of the ${\cal {P}}$-tensor with the Riemann tensor yields\footnote{ As a curious note, one can see that the square of this tensor yields a particular Einstein
plus quadratic gravity in generic $n\ge4$-dimensions 
\[
\text{\ensuremath{{\cal {P}}}}_{\nu\mu\beta\sigma}^{2}=\chi_{\text{GB}}+(n-3)\left(4R_{\mu\nu}^{2}+\frac{R^{2}}{2}(n-6)-2\Lambda\frac{(n-3)}{n-1}\left((n-2)R-\Lambda n\right)\right),
\]
which is not the Lagrangian of critical gravity \cite{Pope,Tahsin_critical} } 
\begin{equation}
R^{\nu\mu\beta\sigma}\text{\ensuremath{{\cal {P}}}}_{\nu\mu\beta\sigma}=\chi_{\text{GB}}-2\Lambda\frac{(n-3)}{n-1}R,
\end{equation}
which vanishes in three dimensions, but gives the Einstein-Gauss-Bonnet
Lagrangian (with a fixed relative coefficient) in generic $n$ dimensions.
The natural question is how one arrives at the ${\cal {P}}$-tensor
(\ref{eq:Ptensor}). We have found the ${\cal {P}}$-tensor from the
following construction: starting from the Bianchi identity 
\begin{equation}
\nabla_{\nu}R_{\sigma\beta\mu\rho}+\nabla_{\sigma}R_{\beta\nu\mu\rho}+\nabla_{\beta}R_{\nu\sigma\mu\rho}=0
\end{equation}
and carrying out the $g^{\nu\rho}$ multiplication, one arrives at
the $\text{\ensuremath{{\cal {P}}}}^{\nu}\thinspace_{\mu\beta\sigma}$
as given in (\ref{eq:Ptensor}) after making use of $\nabla_{\mu}\text{\ensuremath{{\cal {G}}}}^{\mu\nu}=0$ and
$\nabla_{\mu}g_{\alpha\beta}=0$. Note that this still leaves an ambiguity
in the ${\cal {P}}$-tensor, since one can add an arbitrary constant
times $g^{\mu\sigma}g^{\beta\nu}$, but that part can be fixed by demanding
that the ${\cal {P}}$-tensor has the symmetries of the Riemann tensor
and also vanishes for constant curvature backgrounds, which we assumed.
This tensor turns out to be extremely useful in finding conserved
charges of Einstein's gravity for asymptotically $AdS$ spacetimes
for $n>3$ dimensions. Recently in \cite{newformulation}, we gave a brief account of this formulation in Einstein's theory, and in the current work, we shall extend this formulation
to quadratic and generic gravity theories.

The main motivation of our construction is the following: outside
the localized sources, the properties of gravity are fully encoded
in the Riemann tensor. One would naturally expect that the charge
expression, which is an integral on the boundary of a spacelike surface,
would also involve the Riemann tensor at infinity. But a straightforward
computation shows that this is not the case, as we shall revisit in
the next section. The existing formulas involve the first derivatives
of the metric perturbation. The crux of the matter is that the existing
expressions are based on conserved currents which are only gauge-invariant up to a boundary term that vanishes. Our formalism remedies this and constructs an explicitly gauge-invariant current and simplifies the charge expressions significantly.

The layout of the paper is as follows: In section II, which is the
bulk of the paper, we discuss the conserved Killing charges in generic
gravity and give a compact expression that utilizes the ${\cal {P}}$-tensor.
In section III, we discuss the gauge-invariance issue of the conserved
currents. In section IV, we study the $n$-dimensional Schwarzschild-$AdS$
spacetime and the $AdS$ soliton. In \cite{newformulation}, we studied the Kerr-$AdS$ solution an hence we shall not repeat it here. 

\section{{\normalsize{}{}Conserved Charges}}

Conserved charges of generic gravity theory in asymptotically $AdS$
spacetimes were constructed in \cite{Deser_Tekin} as an extension
of the Abbott-Deser charges \cite{Abbott_Deser} of the cosmological Einstein
theory. The latter is a generalization of the ADM charges \cite{ADM}
which are valid for asymptotically flat spacetimes. A detailed account
of these constructions was recently given in \cite{Adami} and for
related constructions, see \cite{Petrov} and \cite{Compere}. Here,
for the sake of completeness, we will briefly summarize the salient
parts of this construction. Consider a generic gravity theory defined
by the field equations depending on the Riemann tensor (${\cal {R}}$),
its derivatives, and contractions 
\begin{equation}
\mathcal{E}_{\mu\nu}\left(g,{\cal {R}},\nabla{\cal {R}},{\cal {R}}^{2},...\right)=\kappa\tau_{\mu\nu},\label{eq:EoM}
\end{equation}
where $\nabla_{\mu}\mathcal{E}^{\mu\nu}=0$, and $\kappa$ is the
$n$-dimensional Newton constant while $\tau_{\mu\nu}$ represents
a localized conserved source. A nontrivial, partially conserved current
arises after one splits the metric as 
\begin{equation}
g_{\mu\nu}=\bar{g}_{\mu\nu}+\kappa h_{\mu\nu},\label{eq:Metric_decomp}
\end{equation}
which yields a splitting of the field equations as 
\begin{equation}
\kappa(\mathcal{E}_{\mu\nu})^{\left(1\right)}\left(h\right)=\kappa\tau_{\mu\nu}-\kappa^{2}(\mathcal{E}_{\mu\nu})^{\left(2\right)}\left(h\right)+{\cal {O}}\left(\kappa^{3}\right),
\end{equation}
where we assumed that $\bar{g}$ solves the field equations, $\mathcal{E}_{\mu\nu}\left(\bar{g}\right)=0$,
exactly in the absence of any source $\tau_{\mu\nu}$ and $(\mathcal{E}_{\mu\nu})^{\left(1\right)}(h):=\frac{d}{d\kappa}\mathcal{E}_{\mu\nu}(\bar{g}+\kappa h)\mid_{\kappa=0}$.
Hence defining $(\mathcal{E}_{\mu\nu})^{\left(1\right)}:=T_{\mu\nu}$,
one has the desired partially conserved current, if the background
admits a Killing vector $\bar{\xi}$: 
\begin{equation}
\mathcal{J}^{\mu}:=\sqrt{-\bar{g}}\thinspace\bar{\xi}_{\nu}\,(\mathcal{E}^{\mu\nu})^{\left(1\right)}.
\end{equation}
As usual, making use of the Stokes theorem, given a spacelike hypersurface
$\bar{\Sigma}$, one has the conserved charge for each background Killing vector
\begin{equation}
Q\left(\bar{\xi}\right):=\int_{\bar{\Sigma}}d^{n-1}y\,\sqrt{\bar{\gamma}}\thinspace\bar{n}_{\mu}\,\bar{\xi}_{\nu}\,(\mathcal{E}^{\mu\nu})^{\left(1\right)},\label{charge_genel}
\end{equation}
where we assumed the that $\mathcal{J}^{\mu}$ vanishes at spacelike
infinity. To proceed further and reduce this integral over $\bar{\Sigma}$
to an integral over the boundary $\partial\bar{\Sigma}$, one must
know the field equations and express $\bar{\xi}_{\nu}\,(\mathcal{E}^{\mu\nu})^{\left(1\right)}$
as a divergence of an antisymmetric two tensor. Recently \cite{newformulation},
we have shown that, using the ${\cal {P}}$-tensor of the previous
section, one can reformulate this problem in the cosmological Einstein
theory in $AdS$ spacetimes {\it without} using the explicit form of the
linearized cosmological Einstein tensor. This is possible because
in Einstein spaces (that are not Ricci-flat such as the $AdS$), one has
the nice property that the Killing vector can be derived from an antisymmetric
"potential" $\bar{{\cal {F}}}_{\mu\nu}$ as 
\begin{equation}
\bar{\xi}^{\mu}=\bar{\nabla}_{\nu}\bar{{\cal {F}}}^{\nu\mu},
\end{equation}
where $\bar{{\cal {F}}}^{\nu\mu}=-\frac{2}{\bar{R}}\bar{\nabla}^{\nu}\bar{\xi}^{\mu}$
with $\bar{R}$ being the constant scalar curvature. Although this
result is valid for {\it any} Einstein space as a background, for concreteness, we shall work
in the $AdS$ background, for which we have 
\begin{equation}
\bar{R}_{\mu\alpha\nu\beta}=\frac{2 \Lambda}{\left(n-2\right)\left(n-1\right)}\left(\bar{g}_{\mu\nu}\bar{g}_{\alpha\beta}-\bar{g}_{\mu\beta}\bar{g}_{\alpha\nu}\right),\qquad\bar{R}_{\mu\nu}=\frac{2 \Lambda}{n-2}\bar{g}_{\mu\nu},\qquad\bar{R}=\frac{2n\Lambda}{n-2}.\label{eq:Max_sym_background}
\end{equation}
To find the conserved charges of a gravity theory defined on an asymptotically
$AdS$ spacetime $\mathscr{M}$, let us assume that there is an antisymmetric
two form, $\ensuremath{{\cal {F}}}_{\mu\nu}$, on the manifold. Then
one has the $exact$ equation for any smooth metric 
\begin{equation}
\nabla_{\nu}(\text{\ensuremath{{\cal {F}}}}_{\beta\sigma}\text{\ensuremath{{\cal {P}}}}^{\nu\mu\beta\sigma})-\text{\ensuremath{{\cal {P}}}}^{\nu\mu\beta\sigma}\nabla_{\nu}\text{\ensuremath{{\cal {F}}}}_{\beta\sigma}=0.\label{eq:ddimensionalmainequation}
\end{equation}
Linearization of (\ref{eq:ddimensionalmainequation}) about the $AdS$
background yields 
\begin{equation}
\bar{\nabla_{\nu}}\biggl((\text{\ensuremath{{\cal {P}}}}^{\nu\mu\beta\sigma})^{\left(1\right)}\bar{\text{\ensuremath{{\cal {F}}}}}_{\beta\sigma}\biggr)-(\text{\ensuremath{{\cal {P}}}}^{\nu\mu\beta\sigma})^{\left(1\right)}\bar{\nabla}_{\nu}\bar{\text{\ensuremath{{\cal {F}}}}}_{\beta\sigma}=0,\label{eq:ddimensionalmainequationlinear}
\end{equation}
which is the main equation from which we will read the conserved current.

\subsection{{\normalsize{}{}Einstein's Theory}}

Let us recapitulate the main points of \cite{newformulation}. Using
the following equivalent form of the $\text{\ensuremath{{\cal {P}}}}$-tensor,
written in terms of the cosmological Einstein tensor, 
\begin{equation}
\text{\ensuremath{{\cal {P}}}}^{\nu}\thinspace_{\mu\beta\sigma}:=R^{\nu}\thinspace_{\mu\beta\sigma}+\delta_{\sigma}^{\nu}\text{\ensuremath{{\cal {G}}}}_{\beta\mu}-\delta_{\beta}^{\nu}\text{\ensuremath{{\cal {G}}}}_{\sigma\mu}+\text{\ensuremath{{\cal {G}}}}_{\sigma}^{\nu}g_{\beta\mu}-\text{\ensuremath{{\cal {G}}}}_{\beta}^{\nu}g_{\sigma\mu}+(\frac{R}{2}-\frac{\Lambda\left(n+1\right)}{n-1})\left(\delta_{\sigma}^{\nu}g_{\beta\mu}-\delta_{\beta}^{\nu}g_{\sigma\mu}\right),
\end{equation}
one arrives at its linearized form 
\begin{eqnarray}
({\cal {P}}^{\nu\mu\beta\sigma})^{\left(1\right)}= &  & (R^{\nu\mu\beta\sigma})^{1}+2(\text{\ensuremath{{\cal {G}}}}^{\mu[\beta})^{(1)}\overline{g}^{\sigma]\nu}+2(\text{\ensuremath{{\cal {G}}}}^{\nu[\sigma})^{(1)}\overline{g}^{\beta]\mu}+(R)^{\left(1\right)}\overline{g}^{\mu[\beta}\overline{g}^{\sigma]\nu}\nonumber \\
 &  & +\frac{4\Lambda}{(n-1)(n-2)}(h^{\mu[\sigma}\overline{g}^{\beta]\nu}+\overline{g}^{\mu[\sigma}{h}^{\beta]\nu}),\label{ktensorlinear}
\end{eqnarray}
where the square brackets denote antisymmetrization with a factor
of 1/2. For the particular antisymmetric background tensor 
\begin{equation}
\bar{\text{\ensuremath{{\cal {F}}}}}_{\alpha\beta}:=\bar{\nabla}_{\alpha}\bar{\xi}_{\beta},
\end{equation}
where $\bar{\xi}_{\beta}$ is an $AdS$ Killing vector, one finds
from (\ref{eq:ddimensionalmainequationlinear}) the following conserved current:
\begin{equation}
\bar{\text{\ensuremath{\xi}}}_{\lambda}(\text{\ensuremath{{\cal {G}}}}^{\lambda\mu})^{\left(1\right)}=\frac{(n-1)(n-2)}{4\Lambda\left(n-3\right)}\bar{\nabla_{\nu}}\biggl((\text{\ensuremath{{\cal {P}}}}^{\nu\mu\beta\sigma})^{\left(1\right)}\bar{\text{\ensuremath{{\cal {F}}}}}_{\beta\sigma}\biggr).\label{eq:finallinearequation}
\end{equation}
Comparing this with the integrand of (\ref{charge_genel}), and using
the Stokes theorem one more time, we find the desired result
\begin{equation}
\boxed{\phantom{\frac{\frac{\xi}{\xi}}{\frac{\xi}{\xi}}}Q\left(\bar{\xi}\right)=\frac{(n-1)(n-2)}{8(n-3)\Lambda G\Omega_{n-2}}\int_{\partial\bar{\Sigma}}d^{n-2}x\,\sqrt{\bar{\gamma}}\,\bar{\epsilon}_{\mu\nu}\left(R^{\nu\mu}\thinspace_{\beta\sigma}\right)^{\left(1\right)}\bar{\text{\ensuremath{{\cal {F}}}}}^{\beta\sigma},\phantom{\frac{\frac{\xi}{\xi}}{\frac{\xi}{\xi}}}}\label{newcharge}
\end{equation}
where $(R^{\nu\mu}\thinspace_{\beta\sigma})^{\left(1\right)}$ is
the linearized part of the Riemann tensor about the $AdS$ background.
Observe that on the boundary $({\cal {P}}^{\nu\mu}\,_{\beta\sigma})^{\left(1\right)}=(R^{\nu\mu}\,_{\beta\sigma})^{\left(1\right)}$,
since the linearized Einstein tensor and the linearized scalar curvature
vanish. The barred quantities refer to the background spacetime $\bar{\mathscr{M}}$
with the boundary $\partial\bar{\mathscr{M}}$. The Killing vector
is $\bar{\xi}^{\sigma}$ from which one defines the antisymmetric
tensor as $\bar{{\cal {F}}}^{\beta\sigma}=\bar{\nabla}^{\beta}\bar{\xi}^{\sigma}$.
The spatial hypersurface $\bar{\Sigma}$ is not equal to $\partial\bar{\mathscr{M}}$;
hence $\bar{\Sigma}$ can have a boundary of its own, that is $\partial\bar{\Sigma}$.
Here the antisymmetric 2-form $\epsilon$ has components $\bar{\epsilon}_{\mu\nu}:=\frac{1}{2}\left(\bar{n}_{\mu}\bar{\sigma}_{\nu}-\bar{n}_{\nu}\bar{\sigma}_{\mu}\right)$,
where $\bar{n}_{\mu}$ is a normal one form on $\partial\bar{\mathscr{M}}$
and $\bar{\sigma}_{\nu}$ is the unit normal one form on $\partial\bar{\Sigma}$
and $\bar{\gamma}$ is the induced metric on the boundary. This is
sufficient for the conserved charges of the cosmological Einstein theory
in $AdS$. But for a generic theory, one must carry out an analogous
computation which is what we do next. But before that,  let us note that for the energy
of the spacetime, we have $\bar{\xi}=\partial_{t}$, and (\ref{newcharge})
becomes 
\begin{equation}
E:=Q\left(\partial_{t}\right)=\frac{(n-1)(n-2)}{2(n-3)\Lambda G\Omega_{n-2}}\int_{\partial\bar{\Sigma}}d^{n-2}x\,\sqrt{\bar{\gamma}}\,\left(R^{rt}\thinspace_{rt}\right)^{\left(1\right)}\bar{\nabla}^{r}\bar{\xi}^{t},\label{energy}
\end{equation}
where $r$ is the radial coordinate and one takes $r\rightarrow\infty$
at the end of the computation. Similarly, for the angular momentum, one can take the Killing vector $\bar{\xi}^\mu = (0, ..., 1,0,...,0)$ and carry out the computation. 

\subsection{{\normalsize{}{}Generic Theory}}

Consider a generic gravity theory which starts with the Einsteinian
part as 
\begin{equation}
\mathcal{E}_{\mu\nu}=\frac{1}{\kappa}\left(R_{\mu\nu}-\frac{1}{2}Rg_{\mu\nu}+\Lambda_{0}g_{\mu\nu}\right)+\sigma E_{\mu\nu}=\tau_{\mu\nu},
\end{equation}
where at this stage all we know about the $E_{\mu\nu}$-tensor is
that it is a symmetric divergence-free tensor (which can come from
an action) and $\sigma$ is a dimensionful parameter. To proceed further,
it is better to recast the equation as 
\begin{equation}
\mathcal{E}_{\mu\nu}=\frac{1}{\kappa}\text{\ensuremath{{\cal {G}}}}_{\mu\nu}+\frac{\Lambda_{0}-\Lambda}{\kappa}g_{\mu\nu}+\sigma E_{\mu\nu}=\tau_{\mu\nu},\label{eq:genericfieldequations}
\end{equation}
whose $(A)dS$ vacua are determined by 
\begin{equation}
\bar{\mathcal{E}}_{\mu\nu}=\frac{\Lambda_{0}-\Lambda}{\kappa}\bar{g}_{\mu\nu}+\sigma\bar{E}_{\mu\nu}=0,
\end{equation}
which in general has many vacua depending on the details of the $\bar{E}_{\mu\nu}$
tensor. We shall assume that $\Lambda$ represents any one of the
viable vacua. To find the conserved charges in this theory, we use
the same procedure as the one in the previous section and define 
\begin{equation}
(\mathcal{E}_{\mu\nu})^{\left(1\right)}=T_{\mu\nu},
\end{equation}
where the right-hand side has all the higher order terms 
\begin{equation}
T_{\mu\nu}=\tau_{\mu\nu}-\kappa(\mathcal{E}_{\mu\nu})^{\left(2\right)}-\kappa^{2}(\mathcal{E}_{\mu\nu})^{\left(3\right)}-...\, .
\end{equation}
So we have the background conserved current 
\begin{equation}
\bar{\nabla}_{\nu}\Bigl(\bar{\xi}_{\mu}(\mathcal{E}^{\mu\nu})^{\left(1\right)}\Bigr)=0,
\end{equation}
and the partially conserved current is $\mathcal{J}^{\nu}=\sqrt{-\bar{g}}\thinspace\bar{\xi}_{\mu}\thinspace(\mathcal{E}^{\mu\nu})^{\left(1\right)}$.
Hence we must compute\footnote{ At this stage, we can search for a modified version of the $\text{\ensuremath{{\cal {P}}}}$-tensor which is conserved and whose trace is the ${\mathcal{E}}$-tensor. One can find such a tensor but it does not have the symmetries of the Riemann tensor anymore, and hence it does not make the ensuing computation any simpler.}
\begin{equation}
\bar{\xi}_{\mu}(\mathcal{E}^{\mu\nu})^{\left(1\right)}=\frac{1}{\kappa}\bar{\xi}_{\mu}(\text{\ensuremath{{\cal {G}}}}^{\mu\nu})^{\left(1\right)}-\frac{\Lambda_{0}-\Lambda}{\kappa}\bar{\xi}_{\mu}h^{\mu\nu}+\sigma\bar{\xi}_{\mu}(E^{\mu\nu})^{\left(1\right)}.
\end{equation}
We have already computed the first part in the previous subsection, and
hence, the new parts are the second and the third terms. But when
the theory is not given, one cannot proceed further from this point.
For this reason, let us consider the quadratic theory as an example
which also covers all the $f(Riemann)$ type theories. The action
of the quadratic theory is 
\begin{equation}
I=\int d^{n}x\sqrt{-g}\left(\frac{1}{\kappa}\left(R-2\Lambda_{0}\right)+\alpha R^{2}+\beta R_{\mu\nu}R^{\mu\nu}+\gamma\left(R_{\mu\nu\rho\sigma}R^{\mu\nu\rho\sigma}-4R_{\mu\nu}R^{\mu\nu}+R^{2}\right)\right),\label{eq:Quad_act}
\end{equation}
and the field equations are \cite{Deser_Tekin} 
\begin{align}
 & \frac{1}{\kappa}\left(R_{\mu\nu}-\frac{1}{2}g_{\mu\nu}R+g_{\mu\nu}\Lambda_{0}\right)+2\alpha R\left(R_{\mu\nu}-\frac{1}{4}g_{\mu\nu}R\right)+\left(2\alpha+\beta\right)\left(g_{\mu\nu}\Box-\nabla_{\mu}\nabla_{\nu}\right)R\nonumber \\
 & +2\gamma\left(RR_{\mu\nu}-2R_{\mu\sigma\nu\rho}R^{\sigma\rho}+R_{\mu\sigma\rho\tau}R_{\nu}^{\phantom{\nu}\sigma\rho\tau}-2R_{\mu\sigma}R_{\phantom{\sigma}\nu}^{\sigma}-\frac{1}{4}g_{\mu\nu}\left(R_{\alpha\beta\rho\sigma}R^{\alpha\beta\rho\sigma}-4R_{\alpha\beta}R^{\alpha\beta}+R^{2}\right)\right)\nonumber \\
 & +\beta\Box\left(R_{\mu\nu}-\frac{1}{2}g_{\mu\nu}R\right)+2\beta\left(R_{\mu\sigma\nu\rho}-\frac{1}{4}g_{\mu\nu}R_{\sigma\rho}\right)R^{\sigma\rho}=\tau_{\mu\nu}.\label{eq:EoM_quad}
\end{align}
Inserting (\ref{eq:Max_sym_background}) in the last equation, one
finds the equation satisfied by $\Lambda$: 
\begin{equation}
\frac{\Lambda-\Lambda_{0}}{2\kappa}+\Biggl((n\alpha+\beta)\frac{(n-4)}{(n-2)^{2}}+\gamma\frac{(n-3)(n-4)}{(n-1)(n-2)}\Bigr)\Lambda^{2}\Biggr)=0.
\end{equation}
Defining the constant 
\begin{equation}
c:=\frac{1}{\kappa}+\frac{4\Lambda n}{n-2}\alpha+\frac{4\Lambda}{n-1}\beta+\frac{4\Lambda\left(n-3\right)\left(n-4\right)}{\left(n-1\right)\left(n-2\right)}\gamma,\label{eq:c}
\end{equation}
one can show that the linearized expressions read 
\begin{align}
\sigma\bar{\xi}_{\nu}(E^{\mu\nu})^{\left(1\right)}-\frac{\Lambda_{0}-\Lambda}{\kappa}\bar{\xi}_{\nu}h^{\mu\nu}= & \left(c-\frac{1}{\kappa}+\frac{4\Lambda}{\left(n-1\right)\left(n-2\right)}\beta\right)\bar{\xi}_{\nu}(\text{\ensuremath{{\cal {G}}}}^{\mu\nu})^{\left(1\right)}\nonumber \\
 & +2\beta\bar{\nabla}_{\alpha}\left(\bar{\xi}_{\nu}\bar{\nabla}^{[\alpha}({\cal G}^{\mu]\nu})^{\left(1\right)}+({\cal G}^{\nu[\alpha})^{\left(1\right)}\bar{\nabla}^{\mu]}\bar{\xi}_{\nu}\right)\nonumber \\
 & +(2\alpha+\beta)\bar{\nabla}_{\alpha}\left(2\bar{\xi}^{[\mu}\bar{\nabla}^{\alpha]}(R)^{\left(1\right)}+(R)^{\left(1\right)}\bar{\nabla}^{\mu}\bar{\xi}^{\alpha}\right),
\end{align}
which then yields the desired result 
\begin{align}
\bar{\xi}_{\nu}(\mathcal{E}^{\mu\nu})^{\left(1\right)}= & \left(c+\frac{4\Lambda}{\left(n-1\right)\left(n-2\right)}\beta\right)\bar{\xi}_{\nu}(\text{\ensuremath{{\cal {G}}}}^{\mu\nu})^{\left(1\right)} \nonumber \\
 & +2\beta\bar{\nabla}_{\alpha}\left(\bar{\xi}_{\nu}\bar{\nabla}^{[\alpha}({\cal G}^{\mu]\nu})^{\left(1\right)}+({\cal G}^{\nu[\alpha})^{\left(1\right)}\bar{\nabla}^{\mu]}\bar{\xi}_{\nu}\right)\nonumber \\
 & +(2\alpha+\beta)\bar{\nabla}_{\alpha}\left(2\bar{\xi}^{[\mu}\bar{\nabla}^{\alpha]}(R)^{\left(1\right)}+(R)^{\left(1\right)}\bar{\nabla}^{\mu}\bar{\xi}^{\alpha}\right).\label{chargedanbironce}
\end{align}
Therefore the conserved charges in quadratic gravity in $(A)dS$ read
as 
\begin{equation}
\boxed{\phantom{\frac{\frac{\xi}{\xi}}{\frac{\xi}{\xi}}}Q\left(\bar{\xi}\right)=\frac{\left(n-1\right)\left(n-2\right)}{4\Lambda\left(n-3\right)}\left(c+\frac{4\Lambda\beta}{\left(n-1\right)\left(n-2\right)}\right)\int_{\partial\bar{\Sigma}}d^{n-2}x\,\sqrt{\bar{\gamma}}\,\bar{\epsilon}_{\mu\nu}\left(R^{\nu\mu}\thinspace_{\beta\sigma}\right)^{\left(1\right)}\bar{\text{\ensuremath{{\cal {F}}}}}^{\beta\sigma}.\phantom{\frac{\frac{\xi}{\xi}}{\frac{\xi}{\xi}}}}\label{charge_quad}
\end{equation}

Observe that for asymptotically $AdS$ spacetimes, the second and
third lines in (\ref{chargedanbironce}) do not contribute. But, if
one tries to generalize the above procedure to asymptotically nonconstant
curvature spacetimes, those parts will also contribute generically.
Therefore, for asymptotically $AdS$ spacetimes, the only difference
between the conserved charges in Einstein's theory (\ref{newcharge})
and the quadratic theory is the numerical factor in (\ref{charge_quad}).
Using the ideas presented in \cite{SST}, the above construction can
be extended to any $f(Riemann)$ theory, where $f$ is a smooth function.
This is because, as far as the energy, vacua and particle contents
are considered, any $f(Riemann)$ theory has an equivalent quadratic
action formulation in which one computes only 3 quantities $f(\bar{R}_{\alpha\beta}^{\mu\nu})$,
$\frac{\partial f}{\partial R_{\alpha\beta}^{\mu\nu}}$, $\frac{\partial^{2}f}{\partial R_{\alpha\beta}^{\mu\nu}R_{\eta\delta}^{\rho\sigma}}$
and their contractions to find the $\kappa$, $\alpha$, $\beta$
and $\gamma$ of the theory to insert in (\ref{charge_quad}).  
As this issue is dealt in \cite{SST} and \cite{Amsel}, we refer the reader to these works. So the crucial part is the Einsteinian part which we have studied in the previous section.

\section{{\normalsize{}{}Gauge-Invariance Issue}}

The problem of the gauge transformations of the charge and the current
that yields the charge is important. Clearly, one expects the charge
to be gauge-invariant in any valid formulation, but the current need
not be. In fact earlier constructions of conserved charges \cite{Abbott_Deser,Deser_Tekin}
used gauge-variant currents which yield gauge-invariant charges. Of
course, for the charges to be gauge-invariant, the noninvariance
of the current should be only up to a boundary term that vanishes on the
boundary. Let us show this in the expression of \cite{Deser_Tekin}
for the cosmological Einstein theory: 
\begin{equation}
2\bar{\xi}_{\nu}({\cal G}^{\mu\nu})^{\left(1\right)}=\bar{\nabla}_{\alpha}{\cal {J}}^{\alpha\mu},\label{equationofj}
\end{equation}
where the antisymmetric current is 
\begin{align}
{\cal {J}}^{\alpha\mu}:= & \bar{\xi}^{\alpha}\bar{\nabla}_{\beta}h^{\mu\beta}-\bar{\xi}^{\mu}\bar{\nabla}_{\beta}h^{\alpha\beta}+\bar{\xi}_{\nu}\bar{\nabla}^{\mu}h^{\alpha\nu}-\bar{\xi}_{\nu}\bar{\nabla}^{\alpha}h^{\mu\nu}+\bar{\xi}^{\mu}\bar{\nabla}^{\alpha}h-\bar{\xi}^{\alpha}\bar{\nabla}^{\mu}h\label{explicitjtensor} \nonumber \\
 & +h^{\mu\nu}\bar{\nabla}^{\alpha}\bar{\xi}_{\nu}-h^{\alpha\nu}\bar{\nabla}^{\mu}\bar{\xi}_{\nu}-h\bar{\nabla}^{\alpha}\bar{\xi}^{\mu}.
\end{align}
Consider an infinitesimal coordinate transformation generated by a
vector field $\zeta$ (not to be confused with the Killing field $\xi$);
one has 
\begin{equation}
\delta_{\zeta}h_{\mu\nu}=\bar{\nabla}_{\mu}\zeta_{\nu}+\bar{\nabla}_{\nu}\zeta_{\mu}=\mathscr{L}_{\zeta}\bar{g}_{\mu\nu},
\label{kirk}
\end{equation}
where $\text{\ensuremath{\mathscr{L}}}_{\zeta}$ denotes the Lie derivative
and hence $\delta_{\zeta}h^{\mu\nu}=-\text{\ensuremath{\mathscr{L}}}_{\zeta}\bar{g}^{\mu\nu}$.
It is easy to see that $\delta_{\zeta}({\cal G}^{\mu\nu})^{\left(1\right)}=\mathscr{L}_{\zeta}{\cal \bar{G}}^{\mu\nu}=0$.
But this only implies from (\ref{equationofj}) that one has  the divergence of the gauge-transformed current to vanish
\begin{equation}
\bar{\nabla}_{\alpha}\delta_{\zeta}{\cal {J}}^{\alpha\mu}=0,
\end{equation}
and hence ${\cal {J}}^{\alpha\mu}$ is not necessarily gauge invariant.
In fact one can show that ${\cal {J}}^{\alpha\mu}$ varies, under
the gauge transformations (\ref{kirk}), as 
\begin{equation}
\delta_{\zeta}{\cal {J}}^{\alpha\mu}=\bar{\nabla}_{\nu}\left(\bar{\xi}^{\alpha}\bar{\nabla}^{\nu}\zeta^{\mu}+\bar{\xi}^{\mu}\bar{\nabla}^{\alpha}\zeta^{\nu}+\bar{\xi}^{\nu}\bar{\nabla}^{\mu}\zeta^{\alpha}+2\zeta^{\alpha}\bar{\nabla}^{\nu}\bar{\xi}^{\mu}+\zeta^{\nu}\bar{\nabla}^{\mu}\bar{\xi}^{\alpha}-(\mu\leftrightarrow\alpha)\right).
\end{equation}
Clearly, since the variation is a boundary term and since ${\cal {J}}^{\alpha\mu}$
is the integrand on the boundary of the spatial slice, the boundary
term does not contribute to the charges (as $\partial\partial\bar{\Sigma}=0$)
and hence the charge is gauge invariant. But this exercise shows us
that the current (\ref{explicitjtensor}) is only gauge invariant
up to a boundary term. 

On the other hand, since $\delta_{\zeta}(R^{\nu\mu}\thinspace_{\beta\sigma})^{\left(1\right)}$
is gauge invariant, our charge expression (\ref{newcharge}) is explicitly
gauge invariant without an additional boundary term. Let us show this:
\begin{equation}
\delta_{\zeta}(R^{\nu\mu}\thinspace_{\beta\sigma})^{\left(1\right)}=\bar{g}^{\alpha\mu}\delta_{\zeta}(R^{\nu}\thinspace_{\alpha\beta\sigma})^{\left(1\right)}-\bar{R}^{\nu}\thinspace_{\alpha\beta\sigma}\delta_{\zeta}h^{\alpha\mu}.\label{gaugeinvarianceriemannuudd}
\end{equation}
Given the linearized Riemann tensor as 
\begin{equation}
(R^{\nu}\thinspace_{\alpha\beta\sigma})^{\left(1\right)}=\bar{\nabla}_{\beta}(\Gamma_{\sigma\alpha}^{\nu})^{\left(1\right)}-\bar{\nabla}_{\sigma}(\Gamma_{\beta\alpha}^{\nu})^{\left(1\right)},\label{eq:firstorderriemann}
\end{equation}
one needs 
\begin{equation}
\delta_{\zeta}(\Gamma_{\sigma\alpha}^{\nu})^{\left(1\right)}=\bar{\nabla}_{\sigma}\bar{\nabla}_{\alpha}\zeta^{\nu}+\bar{R}^{\nu}\thinspace_{\alpha\rho\sigma}\zeta^{\rho}.
\end{equation}
Collecting all the pieces together, one arrives at 
\begin{equation}
\delta_{\zeta}(R^{\nu\mu}\thinspace_{\beta\sigma})^{\left(1\right)}=\text{\ensuremath{\mathscr{L}}}_{\zeta}\bar{R}^{\nu\mu}\thinspace_{\beta\sigma}.
\end{equation}
For the AdS background, one clearly has $\text{\ensuremath{\mathscr{L}}}_{\zeta}\bar{R}^{\nu\mu}\thinspace_{\beta\sigma}=0$
and hence $\delta_{\zeta}(R^{\nu\mu}\thinspace_{\beta\sigma})^{\left(1\right)}=0$,
and so $\delta_{\zeta}Q=0$ as expected. So in our formalism, not
only is the charge explicitly gauge-invariant, but also the current
is explicitly gauge invariant.

In addition to the above discussion of gauge invariance which amounts to changing the coordinates under which the field transforms as (\ref{kirk}), one can consider transformations which are {\it{isometries}} of the background spacetime. Under the latter transformations, the $h_{\mu \nu}$ field transforms as a $(0,2)$ tensor field. In the case of AdS spacetime, these transformations form the group $O(D-1,2)$ for $D \ge 4$ and an infinite dimensional group in $D=3$ dimensions \cite{Brown_Henneaux}. As opposed to the "gauge symmetries " above, these are genuine symmetries of the background spacetime.  Namely, the generators of these transformations are conserved non-trivial charges.  Of course, the components of the background Killing vectors transform as vectors under these isometries, and in general, one gets a superposition of Killing vectors which is a Killing vector itself.  So, as expected, the conserved charges (as generators of symmetries) satisfy the isometry algebra. 

\section{{\normalsize{}{} Some asymptotically $AdS$ spacetimes}}

We have given the computation of the energy and the angular momentum
of the four dimensional Kerr-$AdS$ solution in \cite{newformulation};
here, let us give two more examples.

\subsection{{\normalsize{}{}$n$-dimensional $AdS$-Schwarzschild spacetime}}

Consider a spherically symmetric metric in $n$-dimensions: 
\begin{equation}
{\rm d}s^{2}=-f(r){\rm d}t^{2}+\frac{1}{f(r)}{\rm d}r^{2}+r^{2}{\rm d}\Omega_{n-2}.\label{ilkmetrik}
\end{equation}
For the choice 
\begin{equation}
f(r)=1-\left(\frac{r_{0}}{r}\right)^{n-3}+\frac{r^{2}}{\ell^{2}},\hskip1cm\ell^{2}\equiv-\frac{(n-1)(n-2)}{2\Lambda},\label{f1}
\end{equation}
the metric (\ref{ilkmetrik}) is an Einstein spacetime with the Ricci
tensor 
\begin{equation}
R_{\mu\nu}=-\frac{n-1}{\ell^{2}}g_{\mu\nu}.
\end{equation}
As such, it solves the $n$-dimensional vacuum Einstein's equations
with a cosmological constant. For the Killing vector $\xi^{\mu}=(-1,0,...,0)$,
one needs to compute the integrand in (\ref{energy}) which boils
down to computing the expression $(R^{rt}\,_{rt})^{(1)}\bar{\nabla}^{r}\bar{\xi}^{t}$.
It is easy to see that the relevant component of the full Riemann
tensor reads 
\begin{equation}
R^{rt}\,_{rt}=-\frac{1}{2}f''(r),\hskip1cm(R^{rt}\,_{rt})^{(1)}=-(n-3)(n-2)\frac{r_{0}^{n-3}}{r^{n-2}},
\end{equation}
where $f'(r)=\frac{df(r)}{dr}$. Similarly, one has $\bar{\nabla}^{r}\bar{\xi}^{t}=-\frac{f'}{2}|_{r_{0}\rightarrow0}=-\frac{r}{\ell^{2}}$.
Combining all these in (\ref{energy}) for $G=1$, one finds
the energy of (\ref{ilkmetrik}) as 
\begin{equation}
E=\frac{n-2}{4}r_{0}^{n-3},
\end{equation}
which is exactly the one computed in \cite{Deser_Tekin}. In four dimensions, one has $r_0 = 2 m$ and $E = m$. 

\subsection{{\normalsize{}{} $AdS$ Soliton}}

The metric of the "$AdS$ Soliton" was found by Horowitz-Myers \cite{horowitz}
and reads as 
\begin{eqnarray}
ds^{2}=\frac{r^{2}}{\ell^{2}}\left[\left(1-\frac{r_{0}^{p+1}}{r^{p+1}}\right)d\tau^{2}+\sum_{i=1}^{p-1}(dx^{i})^{2}-dt^{2}\right]+\left(1-\frac{r_{0}^{p+1}}{r^{p+1}}\right)^{-1}\,\frac{\ell^{2}}{r^{2}}\,dr^{2}.\label{adssoliton}
\end{eqnarray}
We shall not go into the physical meaning of this solution, which is
obtained from a $p$-brane metric: the Cartesian coordinates $x^{i}$
($i=1,...,p-1$) and the $t$ denote the coordinates on the ``brane"
and $r\ge r_{0}$. The solution does not have a singularity if the coordinate
$\tau$ is periodic with a period $\beta=4\pi\ell^{2}/(r_{0}(p+1))$.
Consider the timelike Killing vector 
\begin{equation}
\bar{\xi}^{\mu}=(-1,0,...,0),
\end{equation}
then, $\bar{\nabla}^{r}\bar{\xi}^{t}=\frac{-r}{\ell^{2}}$. The relevant
linearized Riemann tensor component can be computed to be 
\begin{equation}
(R^{rt}\,_{rt})^{(1)}=-\frac{(n-3)}{2\ell^{2}}\frac{r_{0}^{n-1}}{r^{n-1}},
\end{equation}
which also shows that there is no $n=3$ $AdS$ soliton. Making use
of (\ref{energy}), one obtains 
\begin{equation}
E=-\frac{V_{n-3}\,\pi}{(n-1)\,\Omega_{n-2}}\,\frac{r_{0}^{n-2}}{\ell^{n-2}}\,,
\end{equation}
where $V_{n-3}$ is the volume of the compact dimensions. This result
matches the one obtained in \cite{Cebeci}.

\section{{\normalsize{}{}Conclusions}}

In a gauge or gravity theory, the conserved charges make sense if
they are gauge or coodinate invariant (at least for small transformations).
The ADM \cite{ADM} and AD \cite{Abbott_Deser} charges and their
generalizations to higher order gravity \cite{Deser_Tekin}, are all
gauge invariant. Namely, they are invariant under small diffeomorphisms. 
[Large diffeomorphisms are a different story; even the flat Minkowski space, while remaining flat, 
can be assigned any mass value in a coordinate system that does not have proper asymptotics.  
See \cite{Adami} for a brief review of this issue.] However, the \textit{explicit}
expressions of these charges do not involve the relevant gauge-invariant
quantity, that is, the linearized Riemann tensor with two up and two
down indices, $(R^{\mu\nu}\,_{\sigma\rho})^{(1)}$, but instead they
involve the first covariant derivative of the metric perturbation
as $\bar{\nabla}_{\alpha}h_{\mu\nu}$ contracted with the Killing
vector in such a way that the final result is gauge invariant \textit{only}
up to a divergence term which vanishes on the boundary. The obvious
question is to try to understand if the gauge-invariant charges can
be written in an \textit{explicitly} gauge-invariant way with the
help of the Riemann tensor. 

There is a stronger motivation for such
a search: outside the sources, the Riemann tensor carries all the
information about gravity. Naturally, it must carry the information
about the conserved charges. It turns out, as we have shown recently
\cite{newformulation} and here, this is indeed the case and the conserved
charge is basically a flux of the Riemann tensor at spatial infinity
contracted with an antisymmetric 2-tensor. The construction is somewhat
nontrivial and is valid only for asymptotically $AdS$ spacetimes 
(which can be generalized to Einstein spacetimes with a non-zero scalar
curvature). For generic gravity theories, the construction is analogous, but there arise many more terms in the final expressions. For an example of this, see the quadratic gravity studied in \cite{Devecioglu}.

The reason that one can write the conserved charges as a flux of the
linearized Riemann tensor at all is that for $AdS$ spacetimes a given Killing
vector $\bar{\xi}^{\mu}$ has an antisymmetric 2-form potential
as $\bar{\xi}^{\mu}=\bar{\nabla}_{\nu}\bar{{\cal {F}}}^{\mu\nu}$,
which helps bring another covariant derivative in the conserved charges
whenever the Killing vector appears, converting the expression to
the linearized Riemann tensor that has two covariant derivatives of
the metric perturbation. To find the charge expression, we used a divergence-free rank-4
tensor whose trace is the Einstein tensor. Interestingly, this construction
is valid only for $n\ge4$ dimensions and is not valid in three dimensions,
since the Riemann tensor can be expressed directly in terms of the
Einstein tensor in three dimensions, the linearized version of which vanishes at spatial infinity.

Finally, we should note that it was realized a long time ago by Regge and Teitelboim \cite{Regge} that in the fully nonlinear Hamiltonian treatment of general relativity in spatially open manifolds one has to include a boundary term $E[g]$ to the bulk Hamiltonian for the functional derivatives of the functionals with respect to the canonical fields to make sense. Namely, to reproduce Einstein's equations from Hamilton's equations,  one must add a surface term to the Hamiltonian which, on the appearance, does not modify  Hamilton's equations  but makes them well-defined. That term turns out to be the ADM energy ($E[g] = E_{\text{ADM}}[h]$) given by the derivative of linearized metric at spatial infinity. Moreover, the {\it value} of the full Hamiltonian, say $H$, which is a sum of the bulk and boundary terms, yields $H  = E_{ADM}[h]$  upon use of the field equations. Namely, the apparently linear-looking ADM energy captures all the nonlinear energy stored in the gravitational field and the localized matter in the bulk of the spacetime.  The same construction works for the Abbott-Deser energy in asymptotically AdS spacetimes and here we have given an explicitly gauge-invariant formulation of this energy and other conserved charges.

\end{document}